\documentclass[twocolumn,aps,pra,twocolumn]{revtex4-1}
\usepackage[latin9]{inputenc}
\usepackage{mathtools}
\usepackage{amsbsy}
\usepackage{amstext}
\usepackage[unicode=true,pdfusetitle,
 bookmarks=false,
 breaklinks=false,pdfborder={0 0 1},backref=false,colorlinks=false]
 {hyperref}
\hypersetup{
 bookmarksnumbered=false,bookmarksopen=false}

\usepackage{hyperref}
\hypersetup{colorlinks,citecolor=NavyBlue,filecolor=black,linkcolor=black,urlcolor=black}
\usepackage[dvipsnames]{xcolor}
\usepackage{dsfont}

\makeatletter

\@ifundefined{textcolor}{}
{%
 \definecolor{BLACK}{gray}{0}
 \definecolor{WHITE}{gray}{1}
 \definecolor{RED}{rgb}{1,0,0}
 \definecolor{GREEN}{rgb}{0,1,0}
 \definecolor{BLUE}{rgb}{0,0,1}
 \definecolor{CYAN}{cmyk}{1,0,0,0}
 \definecolor{MAGENTA}{cmyk}{0,1,0,0}
 \definecolor{YELLOW}{cmyk}{0,0,1,0}
}

\newcommand\ket[1]{\left|#1\right\rangle}
\newcommand\bra[1]{\left\langle #1 \right|}
\newcommand{\adag}{a^{\dagger}}
\newcommand{\adaga}{a^{\dagger}a}

\newcommand{\more}{more\\more\\more\\more\\more}

\usepackage{amsmath}
\usepackage{graphicx}
\usepackage{amssymb}
\usepackage{txfonts,color}
\makeatother

\begin{document}
\title{Measurement-based cooling of a nonlinear mechanical resonator}
\author{Ricardo Puebla}\email[]{r.puebla@qub.ac.uk} 
\affiliation{Centre for Theoretical Atomic, Molecular and Optical Physics, \\School of Mathematics and Physics, Queen's University Belfast, Belfast BT7 1NN, United Kingdom}
\author{Obinna Abah}\email[]{o.abah@qub.ac.uk} 
\affiliation{Centre for Theoretical Atomic, Molecular and Optical Physics, \\School of Mathematics and Physics, Queen's University Belfast, Belfast BT7 1NN, United Kingdom}
\author{Mauro Paternostro}
\affiliation{Centre for Theoretical Atomic, Molecular and Optical Physics, \\School of Mathematics and Physics, Queen's University Belfast, Belfast BT7 1NN, United Kingdom}

\begin{abstract}
  We propose two measurement-based schemes to cool a nonlinear mechanical resonator down to energies close to that of its ground state.  The protocols rely on projective measurements of a spin degree of freedom, which interacts with the resonator through a Jaynes-Cummings interaction. We show the performance of these cooling schemes, that can be either concatenated -- i.e. built by repeating a sequence of  dynamical evolutions followed by projective measurements -- or single-shot. 
  We characterize the performance of both cooling schemes with numerical simulations, and pinpoint the effects of decoherence and noise mechanisms.  Due to the ubiquity and experimental relevance of the Jaynes-Cummings model, we argue that our results can be applied in a variety of experimental setups.
  \end{abstract}

\maketitle

\section{Introduction}\label{intro}

Cooling quantum systems in a finite time down to their ground state is an essential task for the majority of quantum-based technologies~\cite{King:98,Nielsen,Farhi:01,Ursin:07,Giovannetti:11,Georgescu:14,Degen:17}. Although it is possible to isolate and control a quantum systems, the temperature of its surroundings may  still be too large to prepare its quantum ground state with the desired fidelity. This thus demands the development of cooling protocols to enable the preparation of quantum ground states with unit fidelity, and in a short time to overcome the impact of environmental disturbances. Such cooling schemes typically require a hybrid system comprising of two, or more, interacting systems of both discrete (e.g. atomic) and continuous (e.g. vibrational mode) degrees of freedom. Among different methods, it is worth mentioning Doppler~\cite{Diedrich:89} and resolved-sideband cooling, which can be performed depending on the lifetime of the bosonic mode system and leading to distinct final temperatures (cf. Refs.~\cite{Neuhauser:78,Wineland:78,Diedrich:89,Monroe:95} for the development of these techniques in trapped-ions). 

Over the last decades, different means of achieving motional ground state cooling of nano- and micro-mechanical oscillators have been studied, both theoretically and experimentally (cf. Ref.~\cite{Poot:12,Aspelmeyer:14} and references therein). As in trapped-ions, sideband cooling has been demonstrated in these setups~\cite{Metzger:04,Arcizet:06,Schliesser:08,OConnell:10,Chan:11,Teufel:11}. However, other techniques may offer advantages with respect to the standard sideband cooling. Among them, we can mention bang-bang cooling~\cite{Zhang:05}, control state-swapping cooling~\cite{Wang:11} and measurement-based cooling~\cite{Bergenfeldt:09,Li:11,Vanner:13,Rao:16,Khosla:18,Montenegro:18}, which is also known as stochastic cooling due to the probabilistic nature of quantum measurements~\cite{Eschner:95}. 

In this context, the cooling of mechanical systems is of paramount relevance. Mechanical resonators are important components in many electronic systems, while being widely employed in sensors for mass, force, and fields. Recent advancements in fabrication techniques have made possible the realization of micro- and nano-mechanical resonators with high sensitivity and response frequency~\cite{Roukes:01} (cf. Ref.~\cite{Poot:12} for a review). Interestingly, such push to miniaturization has led to the appearance of nonlinear effects in the dynamic response of such devices, often characterized by multi-stability and hysteresis~\cite{Kozinsky:07,Antonio:12,Yao:13}.  Such nonlinear regime can be accessed or explored in different physical platforms, from trapped ions~\cite{Home:11} to circuit quantum electrodynamics~\cite{Ong:11}, from graphene- and carbon nanotube-based resonators~\cite{Lassagne:09,Eichler:11} to optically trapped nanoparticle~\cite{Gieseler:13}. Recently, they have been observed in a system comprising a nanosphere levitated in a hybrid electro-optical trap~\cite{Fonseca:16}. 

Mechanical nonlinearities can be utilized to enhance energy harvesting via piezoelectric (vibration-to-electricity conversion)~\cite{Cottone:09}, which have a good application potential for solving the challenging issue of energy supply for embedded wireless sensors and portable electromechanical devices~\cite{Wei:17}. In addition, they offer high sensitivity that can be harnessed for signal amplification~\cite{Siddiqi:04}, mass and force sensing~\cite{Aldridge:05} or charge detection~\cite{Krommer:00}.  At the fundamental level, the quantum-to-classical transition, i.e. the exploration of the appearance of quantum effects at a macroscopic scale has been studied in these nonlinear systems~\cite{Peano:04,Katz:07,Katz:08}, where the nonlinearity has been identified as a resource in the generation of nonclassical quantum states~\cite{Lu:15,Albarelli:16,Latmiral:16}. Interesting nonlinearities can be engineered by coupling the mechanical mode to an ancillary finite-dimensional system~\cite{Jacobs:09}, an architecture that can be used to study quantum foundations~\cite{Treutlein:14,Diaz:19}. For example, a setup that consists of a vibrating nanomechanical resonator flux coupled to a superconducting qubit  has been proposed as a testbed for quantum interferometry with massive objects~\cite{Khosla:18}.

In this work we present two protocols to cool a mechanical resonator with a Duffing-type nonlinearity down to its ground state aided by projective measurements performed onto a spin degree of freedom coupled to the resonator via a Jaynes-Cummings interaction term~\cite{Jaynes:63}. 
Our proposals can be carried out with or without radiative decay or polarizing noise acting on the spin, whose effect is crucial in resolved-sideband cooling. Hence, these cooling schemes could be carried out using long-lived spin states, and thus also used for other quantum information processing tasks. In particular, we propose a scheme based on the concatenation of joint time evolution of the bosonic and spin degrees of freedom and projective measurements onto the ground state of the spin. We will refer to this method as concatenated scheme (CS). This method not only improves previous results in ultrafast cooling of a mechanical resonator~\cite{Li:11}, but also shows that the non-Gaussian quantum ground state of nonlinear mechanical resonators can be achieved in a finite-time with a very good fidelity. In addition, we show how to attain ground state cooling upon a single-shot (SS) measurement of the spin. This scheme, although allowing for faster cooling and requiring a smaller number of measurements than its concatenated counterpart, demands a tunable and time-dependent spin frequency. The temporal dependence of the spin frequency can be determined using optimal control techniques, such as chopped-random basis optimization (CRAB)~\cite{Doria:11,Caneva:11,Caneva:11b,vanFrank:16}.  We illustrate the high-quality performance of these two schemes, which are able to bring the thermal occupation number of an initial state of a bosonic mode to values very close to zero even under the presence of distinct decoherence and noise sources. Moreover, as our results rely on the ubiquitous Jaynes-Cummings interacting model between a bosonic and a spin degree of freedom, our results may be applied to different platforms to achieve ground state cooling.

The remainder of this article is organized as follows. In Sec.~\ref{sec2}, we begin by introducing the setup of a nonlinear mechanical resonator coupled to a spin degree of freedom and providing relevant experimental parameters.  In Sec.~\ref{sec3} we present the theoretical scheme to cool the resonator down to its ground state by performing projective measurements onto the spin, either in a repeated/concatenated fashion~(Sec.~\ref{scheme_rep}) or upon a single-shot~(Sec.~\ref{scheme_SS}). We further quantify the non-Gaussianity of the resulting state from the concatenated scheme, Sec.~\ref{scheme_rep}. We provide numerical results supporting the good performance of both methods, Sec.~\ref{scheme_rep} and Sec.~\ref{scheme_SS}. We briefly outline the influence of environment for these two proposed schemes in Sec.~\ref{sec4}.    Finally, we present the main conclusions and outlook in Sec.~\ref{sec5}.


\section{Nonlinear mechanical resonator model}\label{sec2} %
\begin{figure}[t!]
\centering
\includegraphics[width=1.\linewidth]{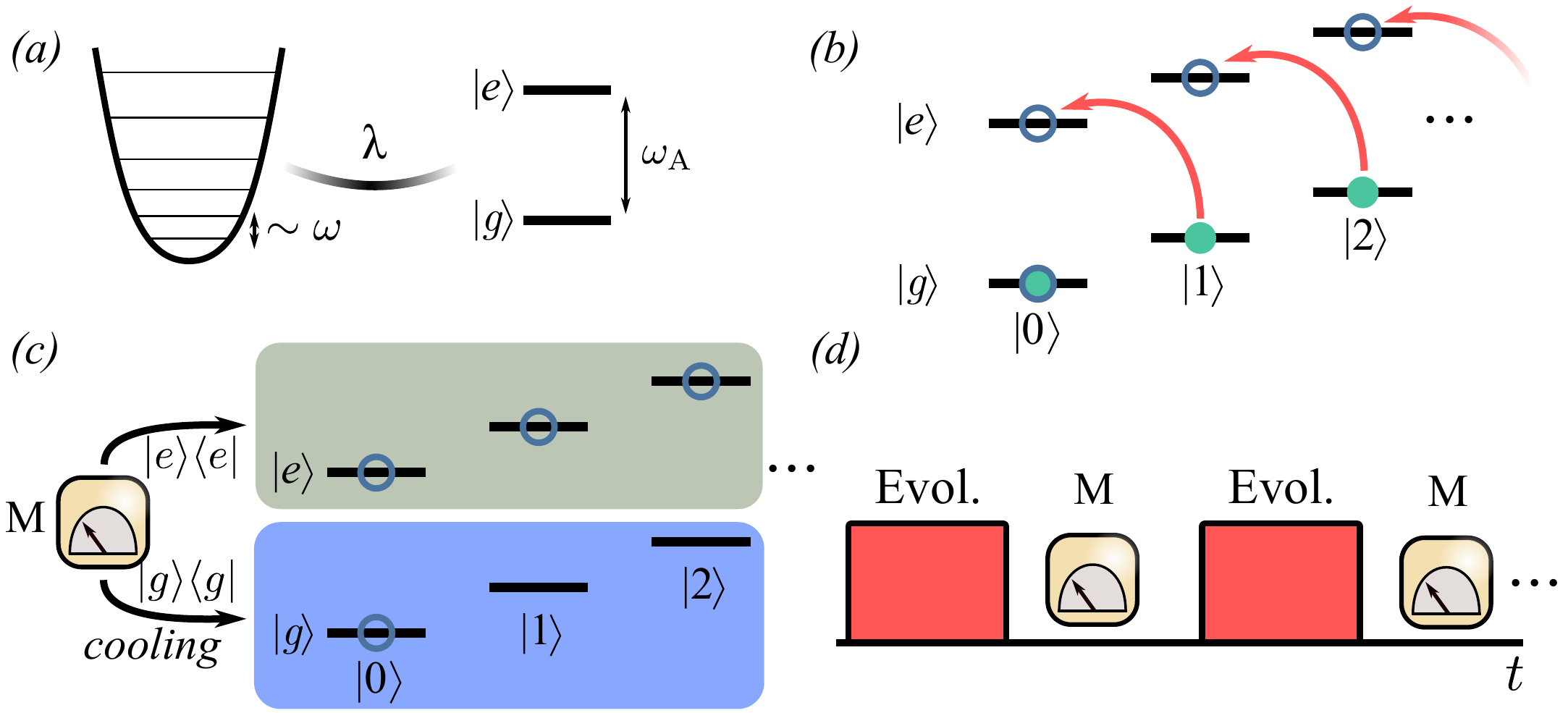}
\caption{\small{(a) Sketch of the non-linear resonator coupled to a spin degree of freedom: The bosonic mode and spin, with frequencies $\omega$ and $\omega_{\rm A}$, respectively, are coupled via a Jaynes-Cummings interaction with strength $\lambda$. (b) Evolution of the initial state, $\rho(0)\!=\!\ket{g}\bra{g}\otimes \rho_b^{\rm th}$ with $\rho_b^{\rm th}\!=\!\sum_{k=0} p_k \ket{k}\bra{k}$ (full circles) and the one sought after the evolution,  which  brings the populations over the Fock states $\ket{n}$ with $n>0$ towards $\ket{e}\bra{e}$ (open circles) while the population of $\ket{g}\bra{g}\otimes\ket{0}\bra{0}$ remains locked, i.e. $\rho(\tau)\!=\!\ket{e}\bra{e}\otimes\sum_{k=0} p_{k+1}\ket{k}\bra{k}+\ket{g}\bra{g}\otimes p_0\ket{0}\bra{0}$.  (c) A projective measurement of the spin onto the eigenbasis of $\sigma_z$. The outcome of applying the projector $M_g\!=\!\ket{g}\bra{g}\otimes \mathds{1}_r$ onto the state $\rho(\tau)$ leaves the bosonic mode in its vacuum state. (d) Cooling scheme by concatenating evolutions plus measurements. }}
\label{fig1}
\end{figure}

Let us consider a bosonic mode of frequency $\omega$, characterized by annihilation and creation operators $a$ and $\adag$, respectively, such that $[a,\adag]\!=\!1$. Such bosonic mode or harmonic oscillator comprises a stiffening Duffing-like deformation with strength $\epsilon>0$, such that $\epsilon\ll\omega$, as found in different experimental platforms. In addition, the bosonic mode is coupled to a (spin-like) two-level system via a Jaynes-Cummings interaction [cf. Fig.~\ref{fig1}(a)]~\cite{Jaynes:63}.  The Hamiltonian of the system reads (we take units such that $\hbar\!=\!1$ throughout the manuscript)
\begin{align}\label{eq:Hs}
H_{\rm s}=\frac{\omega_{\rm A}}{2}\sigma_z+\omega\adaga+\lambda(a\sigma^++\adag \sigma^-)+\frac{\epsilon}{16}(a+\adag)^4,
  \end{align}
where $\omega_{\rm A}$ and $\lambda$ denote the Bohr frequency and coupling strength of the two-level system, respectively. We have introduced the spin Pauli matrices, $\sigma_{x,y,z}$ such that $[\sigma_i,\sigma_j]\!=\!2i\delta_{ijk}\sigma_k$ and $\sigma_{z}\!=\!\ket{e}\bra{e}-\ket{g}\bra{g}$ with $\ket{e}$ ($\ket{g}$) the excited (ground) state of the two-level system. Finally, $\sigma^+=(\sigma^-)^\dag=\ket{e}\bra{g}$ is the spin raising operator. 

The standard Jaynes-Cummings model is recovered by setting $\epsilon\!=\!0$, and thus the ground state of the resonator $H_r\!=\!\omega\adaga+\frac{\epsilon}{16}(a+\adag)^4$ reads as $| \psi_{\rm gs}\rangle\!=\!\ket{0}$ (vacuum) for $\epsilon\!=\!0$ such that $\adaga\ket{n}\!=\!n\ket{n}$, while for $\epsilon/\omega\ll 1$, its ground state can be well approximated by $|\psi_{\rm gs}\rangle\!\approx {\cal N}\left( \ket{0}-3\epsilon/(8\sqrt{2}\omega)\ket{2}-\sqrt{3}\epsilon/(16\sqrt{2}\omega)\ket{4}\right)$, which contains non-zero excitations and is of a non-Gaussian nature~\cite{Teklu:15}. Here, ${\cal N}$ is a normalization constant whose explicit expression is given in Appendix~\ref{app:A}. Hence, as such nonlinear effects are relevant in distinct experimental platforms, the analysis of ground-state cooling based on the occupation number requires a fair comparison with the actual and deformed ground state of the nonlinear resonator. As a result of $\epsilon\neq0$, the number of excitations $N_e\!=\!\adaga+\sigma^+\sigma^-$ is no longer a conserved quantity. However, as we consider a small Duffing perturbation $g,\omega\!\gg\!\epsilon$, the dynamics are mainly governed by the Jaynes-Cummings interaction, i.e. a state  $\ket{g,n+1}$ is transformed into $\ket{e,n}$ at the resonant condition $\omega_{\rm A}\!=\!\omega$ in a time $T_n\!=\!\pi/(2\lambda\sqrt{n+1})$ with $n\geq0$.

Our goal is to cool an initial thermal state of the resonator down to its ground state by performing measurements on the spin degree of freedom (cf. Sec.~\ref{sec3}). That is, the goal consists in performing $\rho^{\rm th}_r\rightarrow |\psi_{\rm gs} \rangle \langle \psi_{\rm gs}|\approx \ket{0}\bra{0}$, with $\rho^{\rm th}_r=\sum_{k=0}p_k\ket{k}\bra{k}$ and $p_k= n_{\rm th}^k/(1+ n_{\rm th})^{k+1}$ where $ n_{\rm th}={\rm Tr}[\adaga \rho_r^{\rm th}]$ is the number of bosonic excitations in the thermal state $\rho_{r}^{\rm th}$. 

The model in Eq.~(\ref{eq:Hs}) 
can be realized  in a number of different platforms. Among them, levitated nanoparticles~\cite{Fonseca:16,Ashley:19}, trapped ions~\cite{Home:11}, circuit quantum electrodynamics~\cite{Ong:11}, optomechanical systems~\cite{Basiri_Esfahani:12,Rips:12,Rips:14}, and cantilever systems~\cite{Katz:08}.  Double-clamped carbon nanotubes can display significant nonlinearities~\cite{Rips:14}: a $\mu$m long carbon nanotube resonator vibrating at $\omega/2\pi\!\approx \!5$ MHz at an environmental temperature of $T_\mathrm{env}\!=\!20$ mK and with a typical quality factor $Q\!\approx\!5\times 10^5$ is endowed with a nonlinear strength $\epsilon/2\pi\!\approx \!200$ KHz ($\epsilon/\omega \sim 4\times 10^{-2}$)~\cite{Rips:12}. Within the optomechanical experimental setup reachable values,  a two-level system defect of frequency $\omega_{\rm A}\in [0.5,1.5]\omega$ coupled to a mechanical resonator, $\omega\approx 200$ MHz, and $Q\approx 10^6$, can achieve spin-boson coupling $\lambda \approx 0.05 \omega$ and spin damping rates $\gamma_d /\omega\in[ 5\times 10^{-8},5\times 10^{-4}]$~\cite{Tian:11}. The amplitude of the resulting Duffing nonlinearity amounts to $\epsilon/\omega\in [ 10^{-4},10^{-5}]$~\cite{Jacobs:09}. For our analysis and without loss of generality, we will choose $\omega_{\rm A}\!\approx\!\omega$, $\lambda\!\lesssim \!0.1\omega$ and scan the values of the ratio $\epsilon/\omega$. The presence of the so-called counter-rotating terms, $\lambda(a^\dagger\sigma^++a\sigma^-)$ which have been neglected in Eq.~\eqref{eq:Hs}, can have a significant impact in the properties of the system~\cite{Casanova:10,Hwang:10,Ashhab:10,Ridolfo:12,Rossatto:17,Kockum:19,Bin:19,Bin:20}, and thus we will discuss its effect on the proposed cooling schemes.


\section{Measurement-based cooling framework}\label{sec3}

We now address the cooling schemes at the core of our proposals 
We study the cooling of a mechanical resonator -- initially prepared in the thermal state $\rho^{\rm th}_r$ -- achieved by combining time-evolution under the total Hamiltonian $H_{\rm s}$ in Eq.~\eqref{eq:Hs}, and projective measurements onto the spin. We consider both the CS and SS approaches, which are described in Sec.~\ref{scheme_rep} and~\ref{scheme_SS}, respectively.

\subsection{Concatenated-measurements scheme}\label{scheme_rep}
Let us now consider the concatenation of $N_{\rm rep}$ time evolutions under the Hamiltonian $H_{s}$ followed by a projective measurement onto the ground state of the spin, described by the projector $M_g$ where $M_{x}\!=\!\ket{x}\bra{x}\otimes \mathds{1}_r$  is the projector onto the spin state $\ket{x}$ and $x\in\{e,g\}$ and $\mathds{1}_r$ is the identity operator acting on the Hilbert space of the resonator. The initial state of the joint system reads 
\begin{align}\label{eq:rho0}
\rho_{s}(t_0=0)=\ket{g}\bra{g}\otimes \rho_{r}^{\rm th}.
  \end{align}
This cooling scheme consists in bringing populations from $\ket{g,n+1}$ to $\ket{e,n}$ states with $n\!\geq\!0$ by sweeping each of the subspaces at a time. This is achieved by evolving $\rho_s(0)$ during a time $T_n\!=\!\pi/(2\lambda\sqrt{n+1})$, i.e., $\rho_s(T_n)\!=\!U(T_n)\rho_s(0)U^{\dagger}(T_n)$ with $U(t)=e^{-i t H_{\rm s}}$ the evolution operator. In this manner, we remove excitations and thus cool down the resonator state by performing a projective measurement $M_g$ on the spin degree of freedom. The state upon the measurement becomes $\rho_s(T_n)\rightarrow M_g\rho_s(T_n)M_g/{\rm Tr}[M_g\rho(T_n)M_g]$. Thus, the state after the first block of evolution and spin measurement is given by
\begin{align}
\rho_s(T_0)=\frac{M_gU(T_0)\rho_s(0) U^{\dagger}(T_0)M_g}{ {\rm Tr}[M_g U(T_0)\rho_s(0) U^{\dagger}(T_0)M_g]},
  \end{align}
where we have chosen $T_0=\pi/(2\lambda)$ as the duration of the first time evolution. This procedure is repeated $N_{\rm rep}$ times, where each repetition comprises a time evolution of duration $T_n=\pi/[2\lambda(n+1)^{1/2}]$ -- with increasing $n$ -- such that the population is transferred from $\ket{g,n+1}$ to $\ket{e,n}$. The total time taken by the cooling process is thus $T_f\!=\!\sum_{n=0}^{N_{\rm rep}-1} T_n=\pi/(2\lambda) \sum_{n=0}^{N_{\rm rep}-1}(n+1)^{-1/2}$, so that $T_f\propto \lambda^{-1}$, and where we have assumed a zero detection time.   The probability of a successful detection of the spin in its ground state $\ket{g}$ upon the evolution $U(T_n)$ is given by $p_{g;n}\!=\!{\rm Tr}[M_g\rho(T_n)M_g]$, which is lower bounded by the probability $p_0=(1+n _{\rm th})^{-1}$ to find the oscillator in its ground state when prepared in the initial thermal state 
$\rho_r^{\rm th}=\sum_{k=0}p_k\ket{k}\bra{k}$ with $p_k\!=\! n_{\rm th}^k/(1+ n _{\rm th})^{k+1}$ and $n_{\rm th}={\rm Tr}[\adaga \rho_r^{\rm th}]$. Upon $N_{\rm rep}$ repetitions, a successful detection probability is given by $p^{\rm sdp}=\Pi_{n=0}^{N_{\rm rep}-1}p_{g;n}$ and $p^{\rm sdp}\approx p_0$ for $N_{\rm rep}\gg 1$. Hence, one can already notice that this method can be favourable to cool down states of a resonator containing few excitations. In particular, if $ n_{\rm th}\lesssim 10$, we have $p_{0}\gtrsim 1/10$ with $p_k\lesssim 10^{-3}$ for $k\gtrsim 50$, so that $N_{\rm rep}\lesssim 50$ would be sufficient to achieve a significant reduction on the occupation number. Recall however that as a consequence of the third law of thermodynamics and the unattainability principle~\cite{Masanes:17}, it is not possible to exactly prepare the ground state of a quantum system in a finite time. Nevertheless, depending on the parameters, the resulting state will be close to the actual ground state. It is worth mentioning that our scheme is similar to the one proposed in Ref.~\cite{Li:11}, although here we do not require random detection times. Indeed, by fixing the evolution times by $T_n$, we boost the cooling performance of the scheme. Before illustrating the performance of this cooling method with numerical simulations, it is worth commenting that depending on the initial thermal occupation $n_{\rm th}$, degree of nonlinearity $\epsilon$ and number of repetitions $N_{\rm rep}$, the final state $\rho_s(T_f)$ will exhibit a large purity and high fidelity with respect to the ground state of the deformed oscillator.

\begin{figure*}[t!]
\centering
\includegraphics[width=1.\linewidth]{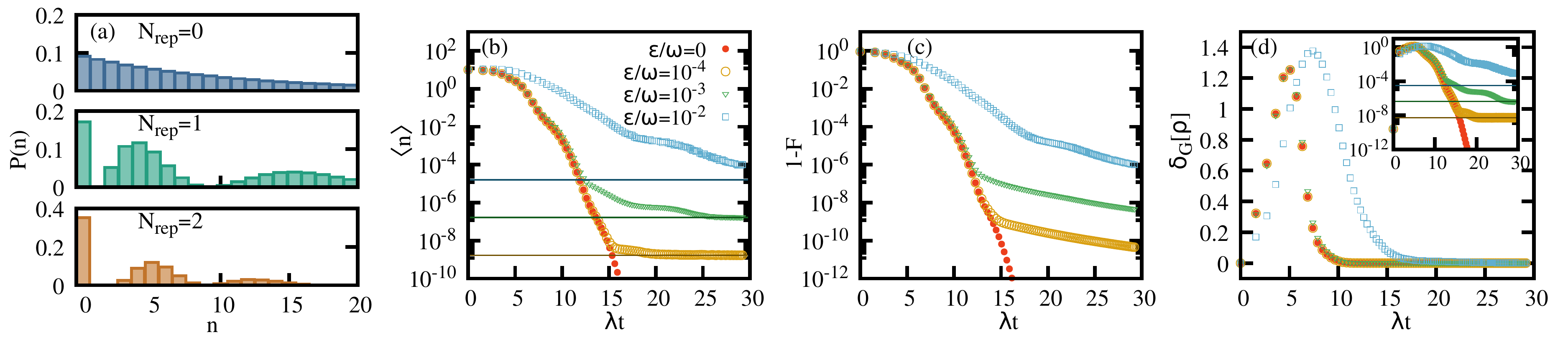}
\caption{{(a) Occupation probability of the $n^\text{th}$ Fock state, $P(n)$ for an initial thermal state with $n_{\rm th}=10$ (top panel), and the resulting distribution after $N_\text{rep}=1$ (middle panel) and $2$ (bottom panel) iterations of the CS for $\epsilon=10^{-4}\omega$ and $\lambda=0.02\omega$. The growth of the ground-state population $P(n=0)$ achieved with more repetitions of the CS is well visible. In (b) we show the average occupation number $\langle n \rangle$ against the dimensionless time $\lambda t$ for $\omega_A=\omega$, an initial state with $n_{\rm th}=10$, and $\epsilon/\omega=0$ (solid circles), $10^{-4}$ (open circles), $10^{-3}$ (open triangles), and $10^{-2}$ (open squares). The solid lines denote the occupation number of the actual ground state of the nonlinear resonator. Panels (c) shows the temporal behavior of the state infidelity $1-F(t)$ with $F(t)=\langle \psi_{\rm gs}| \rho_r(t)|\psi_{\rm gs}\rangle$, which quantifies the distance between the reduced state of the resonator $\rho_r(t)$ at time $t$ and the ground state $|\psi_\text{gs}\rangle$ of the corresponding nonlinear model. (d) Measure of non-Gaussianity $\delta_{\rm G}[\rho_r(t)]$ of the resonator state for the same cases as in panel (b).  
The inset displays the same the plot as in the main panel but in a log-scale for a better illustration, where the solid lines correspond to the degree of non-Gaussianity of $|\psi_{\rm gs}\rangle$.}}
\label{fig2}
\end{figure*}

%
In Fig.~\ref{fig2}(a) we show how the occupation probability of the $n^\text{th}$ Fock state  $P(n)\!=\!\langle n| \rho_r|n \rangle$  changes by performing this protocol. Here, we start with an evolution of duration $T_0$ which brings all the population from $\ket{g,1}$ to $\ket{e,0}$ so that upon the projective measurement onto $\ket{g}\bra{g}$, the population over the Fock state $\ket{1}$ vanishes, i.e. $P(n=1)\!=\!0$. By repeating the process, the vacuum state is achieved with high probability. The average occupation number $\langle n \rangle$ gets largely reduced upon few repetitions, as exemplified in Fig.~\ref{fig2}(b) for an initial state with  $n_{\rm th}\!=\!10$. The ground state of the nonlinear resonator is not $\ket{0}$ for $\epsilon\neq 0$. Our method leads to similar ground-state occupation number, although resonators with large values of $\epsilon$ require longer times to saturate the occupation number. This is due to the nonlinear term in Eq.~\eqref{eq:Hs} (cf. Fig.~\ref{fig2}(b) for $\epsilon/\omega\!=\!10^{-2}$ and $\lambda\!=\!0.02\omega$), which couples different states in the Jaynes-Cummings ladder. The fidelity $F$ of the state $\rho_r$ with respect to the actual ground state of the nonlinear resonator $|\psi_{\rm gs}\rangle$  approaches one, $F\!=\!\langle \psi_{\rm gs}| \rho_r|\psi_{\rm gs}\rangle\approx 1$, upon sufficiently many repetitions (cf. Fig.~\ref{fig2}(c)). The fidelity never reaches one in a finite time, which can be thought of as a consequence of the unattainability principle and the third law of thermodynamics~\cite{Masanes:17}. Nevertheless, the resulting state becomes so close to the actual ground state to display all its features: not only the mean number of excitations in the achieved state is very close to that of the ground state, $\langle n\rangle \approx 21\epsilon^2/128\omega^2$ [cf. Fig.~\ref{fig2}(b) and Appendix~\ref{app:A} for the derivation of the ground-state occupation number], but also other features are accurately reproduced. Here we focus on the degree of non-Gaussianity of the state that we obtain through our protocol. In fact, the nonlinear nature of of the oscillator and the measurement-dependent interaction with the spin result in a pronouncedly non-Gaussian effective dynamics of the mechanical resonator. 
We thus quantify the degree of non-Gaussianity of the state achieved by this cooling scheme using the measure~\cite{Genoni:10} 
\begin{equation}
\delta_G[\rho_r(t)]\!=\!S[\rho_r(t)||\rho_G],
\end{equation}
which is based on the quantum relative entropy of the reduced state $\rho_r(t)$ of the resonator at the generic instant of time $t$ and a reference Gaussian state $\rho_G$ having the same first and second moments of the oscillator's position and momentum operator as $\rho(t)$ (cf. Appendix~\ref{app:B}). 

In Fig.~\ref{fig2}(d) we plot the behavior of $\delta_{\rm G}[\rho_r]$  against the dimensionless time $\lambda t$ and for various choices of the ratio $\epsilon/\omega$. As the initial state of the system is thermal, by definition we have $\delta_{\rm G}[\rho_r(0)]=0$. However, as mentioned above, due to the dynamics the system soon starts developing a non-zero degree of non-Gaussianity,which converges to the value of the ground state $|\psi_{\rm gs}\rangle$ [cf. inset in Fig.~\ref{fig2}(d)]. In between, the dynamics induces a strong non-Gaussian character of $\rho_r(t)$, producing a peak whose location and amplitude depends on the choice of $\epsilon/\omega$. This suggests that, should the goal be that of achieving a state with a large degree of Gaussianity, the protocol can be tailored so as to achieve $\delta_G[\rho_r(t)]\gg\delta_G[|\psi_{\rm gs}\rangle\langle\psi_{\rm gs}|]$, at the cost of a larger occupation number.  

\begin{figure}
\centering
\includegraphics[width=1.\linewidth]{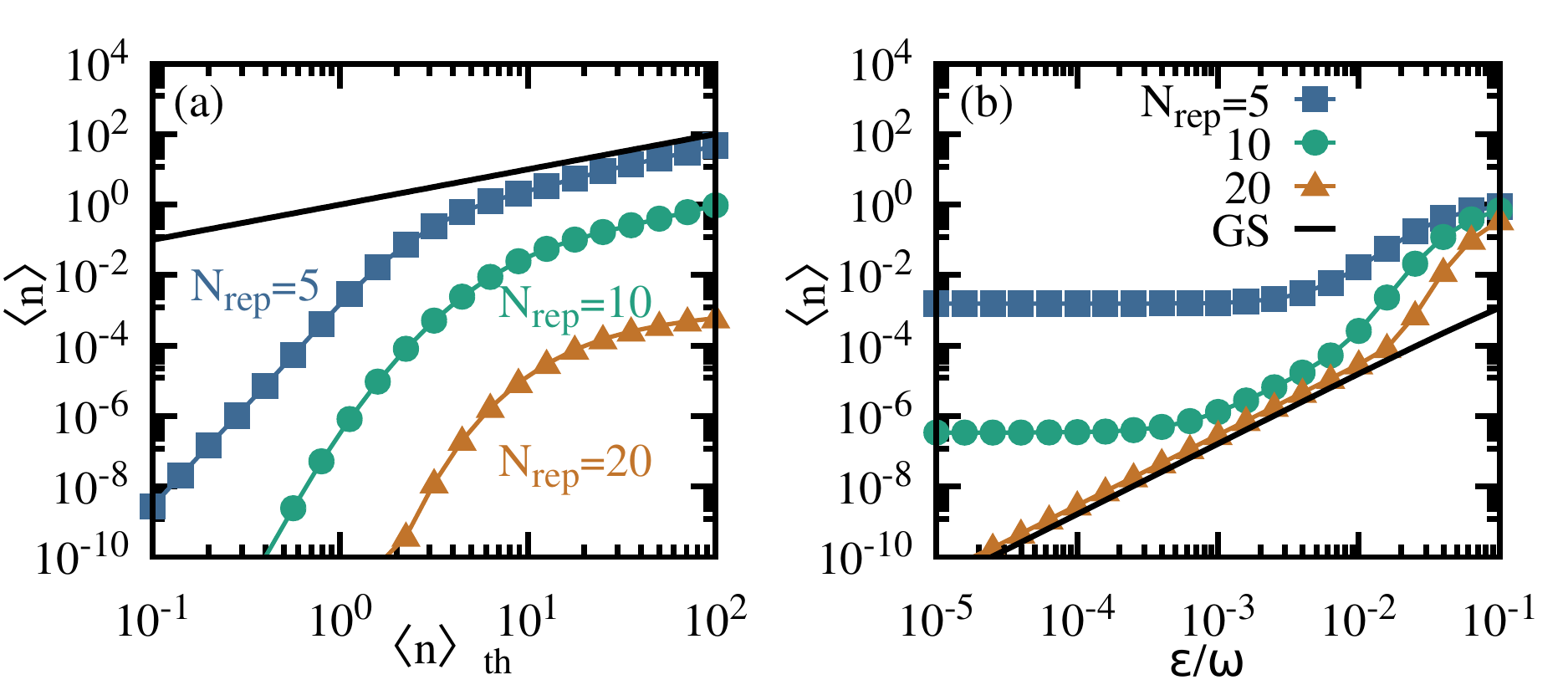}
\caption{\small{(a) Average occupation number $\langle n \rangle$ different $N_{\rm rep}$ repetitions ($5$, $10$ and $20$), as a function of the initial thermal occupation $n_{\rm th}$ for $\epsilon\!=\!0$. The solid black line corresponds to $\langle n \rangle\!=\!n_{\rm th}$ (no cooling). The panel (b) shows the attained $\langle n \rangle$ upon $N_{\rm rep}$ repetitions (again $5$, $10$ and $20$) as a function of the nonlinearity $\epsilon/\omega$ of the resonator, with $n_{\rm th}\!=\!1$. The solid black line denotes the occupation number of the nonlinear ground state, and very close to $\langle n\rangle \approx 21\epsilon^2/128\omega^2$ (cf. Appendix~\ref{app:A}). }}
\label{fig3}
\end{figure}

As commented previously, CS is effective in cooling down thermal states containing $n_{\rm th}\lesssim 10$: larger initial occupation numbers imply a decreasing successful detection probability $p^{\rm sdp}$ and an exceedingly large number of iterations to significantly cool down the state of a nonlinear resonator. This is illustrated in Fig.~\ref{fig3}(a), where the average occupation number $\langle n \rangle$ after $N_{\rm rep}$ repetitions is plotted as a function of the initial thermal occupation $n_{\rm th}$, and for $\epsilon=0$ (chosen as a benchmark case). Indeed, while high temperature states are not so efficiently cooled down with this scheme, states with $n_{\rm th}\lesssim 10$ are brought down to $\langle n \rangle<10^{-4}$ after $N_{\rm rep}\lesssim 20$. The same applies to nonlinear resonators. In Fig.~\ref{fig3}(b) we plot the value of $\langle n \rangle$ achieved after $N_{\rm rep}\!=\!5$, $10$ and $20$ as a function of $\epsilon/\omega$ and for $n_{\rm th}=1$, revealing again that the actual ground state of the nonlinear resonator $|\psi_{\rm gs}\rangle$ can be reached to a very good approximation.

The inclusion of counter-rotating terms in Eq.~\eqref{eq:Hs}, and thus of transitions between $\ket{g,n}\leftrightarrow \ket{e,n+1}$,  may affect the cooling performance depending on the value of $\lambda/\omega$ and the nonlinear contribution $\epsilon/\omega$. For example, for $\epsilon\!=\! 10^{-2}\omega$ and $\lambda\!=\! 0.02\,\omega$, as considered in Fig.~\ref{fig2}, we observe a similar cooling performance. The effect of the counter-rotating terms becomes more evident for $\epsilon/\omega\rightarrow 0$ since $\langle n\rangle\rightarrow 0$, and thus small but non-vanishing transition rates for $\ket{g,n}\leftrightarrow \ket{e,n+1}$ will limit this cooling scheme. Indeed, including counter-rotating terms for $\epsilon\!=\! 10^{-3}\omega$ and $\lambda\!=\! 0.02\omega$ leads to $\langle n\rangle \approx 10^{-4}$ for $\lambda t\approx 30$ (cf. Fig.~\ref{fig2}(b)). We note that the impact of decoherence and dissipation processes, which is discussed in Sec.~\ref{sec4}, will set a tighter constraint on the cooling performance. 


\subsection{Single-shot measurement scheme}\label{scheme_SS}

CS relies on a population transfer from $\ket{g,n+1}$ to $\ket{e,n}$ achieved by sequentially addressing different subspaces with growing $n$. 
In order to overcome the limitation intrinsic to that scheme, we propose an optimal protocol to perform the population transfer $\ket{g,n+1}\rightarrow\ket{e,n}$ for different $n$ simultaneously and in a short time, $\tau\ll T_f\propto N_{\rm rep}^{1/2}\lambda^{-1}$. A single projective measurement $M_g$ at the end of such optimal dynamic protocol will bring the system to its ground state with a very good accuracy. 

In order for the protocol to be effective, though, and to implement the optimal control strategy,  one must allow for a time-dependent parameter to be tuned  externally. In the following we assume that the spin frequency can be controlled in a time-dependent fashion, although similar results can be obtained straightforwardly by selecting another parameter. The initial state $\rho_s(0)$ 
now evolves under the following time dependent Hamiltonian
\begin{align}\label{eq:Hst}
H_{\rm s}(t)=\frac{\omega_{\rm A}(t)}{2}\sigma_z+\omega\adaga+\lambda(a\sigma^++\adag \sigma^-)+\frac{\epsilon}{16}(a+\adag)^4. 
  \end{align}
The shape of the protocol $\omega_{\rm A}(t)$ is then optimized to achieve the desired final state. As an example, consider  $\epsilon\!=\!0$ so that we aim to transform $\rho_s(0)$, as given in Eq.~\eqref{eq:rho0}, into $\rho_s(\tau)\!=\!\ket{g}\bra{g}\otimes p_0\ket{0}\bra{0}+\ket{e}\bra{e}\sum_{k=0} p_{k+1}\ket{k}\bra{k}$, where $\rho_s(\tau)\!=\!U_t(\tau)\rho_s(0)U^{\dagger}_t(\tau)$ and $U_t(\tau)\!=\!\mathcal{T}e^{-i\int_0^\tau dt H_{\rm s}(t)}$ is the time-evolution operator. A single-shot measurement $M_g$ would lead to $\rho_s(\tau)\!=\!\ket{g}\bra{g}\otimes \ket{0}\bra{0}$, i.e., to the ground state of the resonator for $\epsilon\!=\!0$. As in the CS, the success probability of detecting the spin in the state  $\ket{g}$ upon a single shot is lower bounded as $p^{sdp}\geq p_0=1/(1+n_{\rm th})$. 

The optimization is carried out using the technique chopped-random basis approximation (CRAB)~\cite{Doria:11,Caneva:11,Caneva:11b} and a Nelder-Mead search algorithm~\cite{Nelder:65}. Other techniques could be employed equally effectively~\cite{Krotov,Khaneja:05,Watts:15,Bukov:18,Goerz:19}. For convenience, and as $\epsilon/\omega\ll1$, we perform the optimization for $\epsilon=0$, i.e. in a Jaynes-Cummings model, which decouples in a set of Landau-Zener models at different energy spacings (cf. Appendix~\ref{app:C}). We fix $\omega_{\rm A}(0)\!=\!\omega_{\rm A}(\tau)=\omega$, so that the optimization corresponds to finding the coefficients $a_n$ and $b_n$ in
\begin{align}\label{eq:wAt}
\omega_{\rm A}(t)/\omega=1+t\,  (\tau-t)\sum_{n=1}^{N_\omega} \left[a_n\cos(\omega_n t)+b_n\sin(\omega_nt)\right]
\end{align}
where $\omega_n=2\pi n/\tau$ and with a total protocol time $\tau$ longer than the value set by the quantum speed limit~\cite{Deffner:17qsl}. In this case, the minimum time needed to perform such transformation reads as $\tau_{\rm QSL}\equiv T_0\!=\!\pi/(2\lambda)$~\cite{Caneva:11b}. Here we choose $\tau\!=\!3\tau_{\rm QSL}$ although we remark that, provided that $\tau\geq \tau_{\rm QSL}$, an optimal protocol can always be found.  As the achievement of exact ground-state cooling requires the optimization over the infinitely many subspaces of $H_s(t)$, our numerical simulation would lead to the ground state only approximately. 

\begin{figure}[t!]
\centering
\includegraphics[width=1.\linewidth]{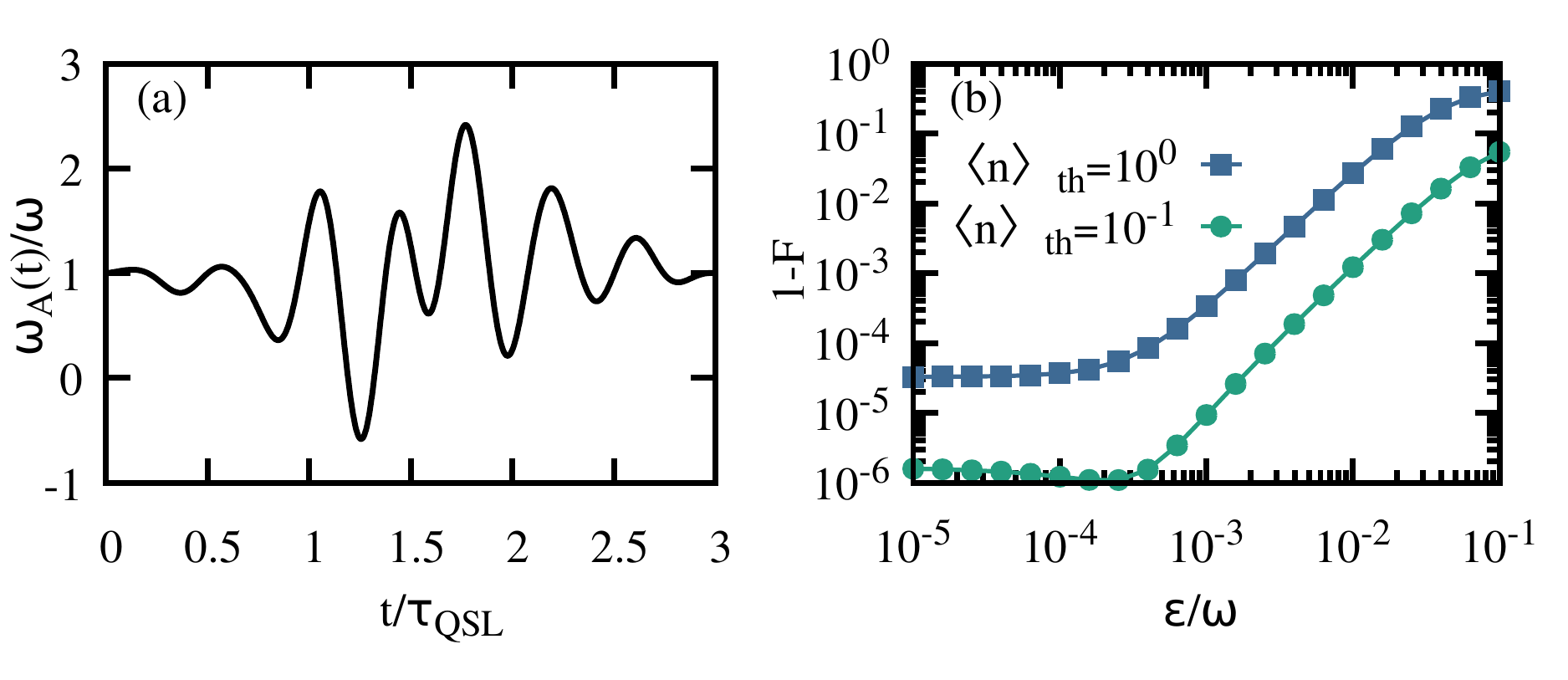}
\caption{\small{(a) An optimized time-dependent spin frequency $\omega_{A}(t)$ found for $N_c=N_\omega=10$ and total time $\tau=3\tau_{\rm QSL}=3\pi/(2\lambda)$. (b) Infidelity of the resulting state upon the time evolution and a projective measurement onto $\ket{g}\bra{g}$ (single-shot) with respect to the ground state of the non-linear resonator, as a function of the parameter $\epsilon$ and for two different initial states, $\langle n\rangle_{th}=1$ and $0.1$. See main text for further details.}}
\label{fig4}
\end{figure}

In Fig.~\ref{fig4}(a) we show a possible optimal form of $\omega_{\rm A}(t)$ obtained by CRAB optimization considering the first $N_c=10$ subspaces of the Jaynes-Cummings model, taking $N_{\omega}=10$ frequencies in Eq.~\eqref{eq:wAt} and $\lambda=\omega/10$. By evolving the initial state Eq.~\eqref{eq:rho0} using such optimal choice, we are able to cool down the non-linear resonator, and get close to its ground state. For $\langle n\rangle_{th}=1$, we find $\langle n\rangle\ll 10^{-2}$ for $\epsilon/\omega <10^{-3}$ with fidelity $F>0.999$ [cf. Fig.~\ref{fig4}(b)]. Finally, it is worth mentioning that the inclusion of the counter-rotating terms in Eq.~\eqref{eq:Hst} can be still carried out via an optimization, although numerically more demanding  as it requires the use of the full Hamiltonian $H_s(t)$.

\begin{figure}[b!]
\centering
\includegraphics[width=1\linewidth]{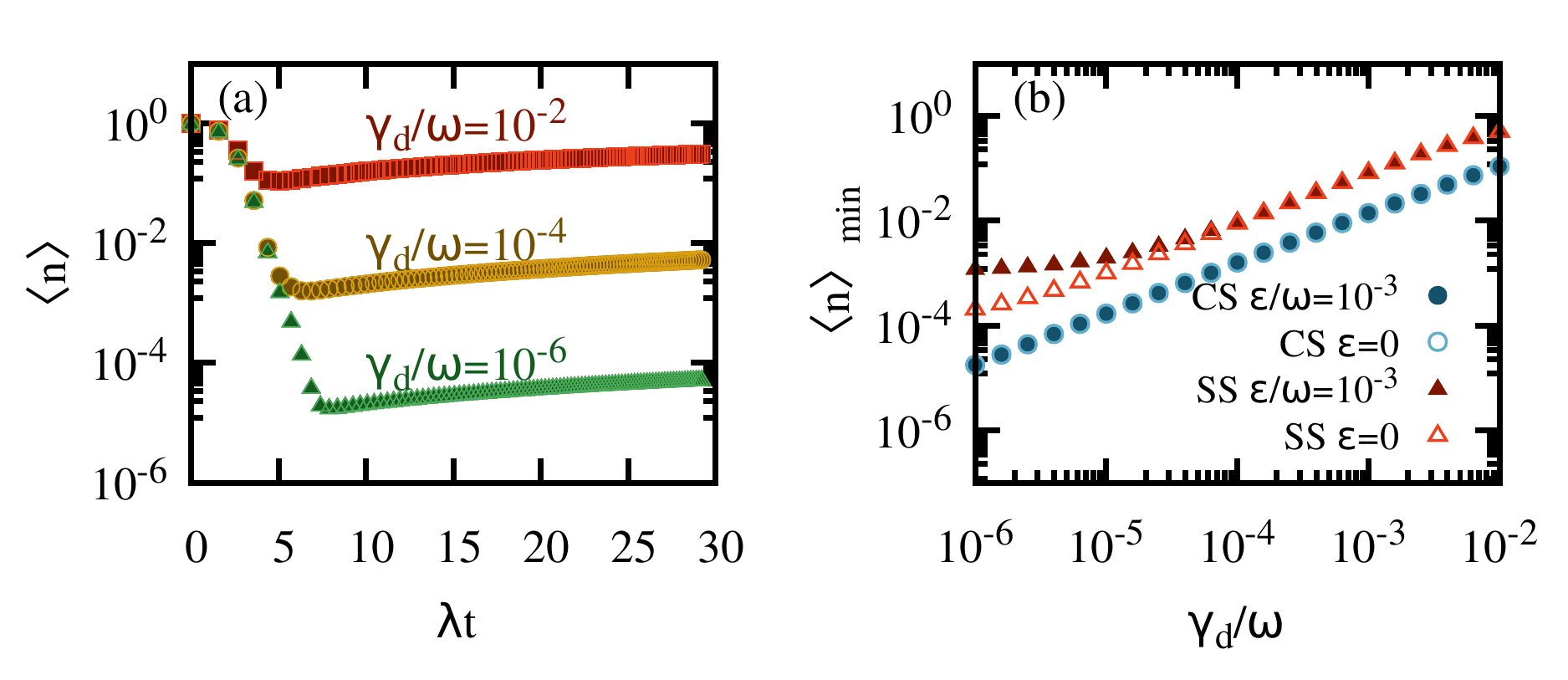}
\caption{\small{(a) Evolution of the average occupation number for the CS for different dissipation rates $\gamma_d$, from top to bottom, $\gamma_d/\omega=10^{-2}$, $10^{-4}$ and $10^{-6}$, and $\epsilon=0$ (open points) and $\epsilon/\omega=10^{-3}$ (full points), which lie on top of each other. In panel (b) we show the minimum value $\langle n\rangle_{\rm min}$ for the CS (circles) as well as the resulting $\langle n\rangle$ upon a SS measurement using the optimized protocol (triangles).}}
\label{fig5}
\end{figure}

\section{Robustness of cooling scheme --- Dynamics in the presence of environmental effects}\label{sec4}

Cooling the resonator down close to its ground state demands an evolution time such that dissipation effects may be significant. We must thus determine the impact of the interaction with an environment on the performance of the protocol. 
Here, we consider the dynamics of the system dictated by the master equation~\cite{Breuer,Rivas}
\begin{align}\label{eq:ME}
\dot{\rho}_r(t)=-i[H_{\rm s},\rho_r(t)]+\mathcal{D}_{a}[\rho_r(t)]+\mathcal{D}_{\adag}[\rho_r(t)],
  \end{align}
where the dissipators have the standard Lindblad form
\begin{align}
  \mathcal{D}_A[\bullet]=\frac{\Gamma_{A}}{2}\left(2A\bullet A^{\dagger}-\{A^{\dagger}A,\bullet \}\right)
\end{align}
with jump operator $A$ and noise rate $\Gamma_A$. In particular, for $a$ and $\adag$, the noise rates are $\Gamma_{a}=\gamma_d(n_{\rm th}+1)$ and $\Gamma_{\adag}=\gamma_d n^{\rm th}$~\cite{Breuer}. Note that for the SS measurement scheme, the dynamics follows from Eq.~\eqref{eq:ME} but with a time-dependent Hamiltonian $H_{\rm s}$. As discussed in Sec.~\ref{sec2}, we consider the experimentally relevant regime $\gamma_d/\omega \in [10^{-6},10^{-2}]$.

In the CS, the average occupation number $\langle n\rangle$ now results a competition between the decreasing behavior due to the cooling scheme and an additional contribution $\langle n\rangle\propto n_{\rm th}(1-e^{-\gamma_d t})$ in the long time limit, due to the dissipation in Eq.~\eqref{eq:ME}~\cite{Breuer}. It is worth stressing that, due to the time-evolution followed by projective measurements, there is a non-trivial interplay between cooling and heating processes. As a result, $\langle n\rangle$ becomes minimal upon a number of repetitions. This is plotted in Fig.~\ref{fig5}(a), where we show the evolution of $\langle n\rangle$ for different parameters $\epsilon$ and $\gamma_d$, with $n_{\rm th}=1$ and $\lambda=\omega/10$. The data points for $\epsilon=0$ and $\epsilon/\omega=10^{-3}$ lie on top each other since the impact of dissipation is stronger that nonlinear effects. The fidelity with respect to the ground state of the nonlinear resonator behaves in a similar manner.

For the SS scheme, one might consider that the system-environment interaction is less relevant as the protocol is performed in a shorter time than in CS. However, in the CS the time between consecutive projective measurements is given by $T_n$, while in the SS the time evolution $\tau$ is  such that  $\tau\geq T_n$, where the equality holds for evolutions performed at the quantum speed limit. The shorter the evolution time, the smaller the impact of the dissipation on the performance of the protocol, and better cooling performance can be achieved. Yet, for $\tau=\tau_{\rm QSL}$, the numerical optimization becomes very demanding. We have analyzed the performance of SS with $\tau>\tau_{\rm QSL}$, finding that dissipation has a larger effect than in CS. In particular, we find that the minimum number of excitations in the resonator during the application of CS, $\langle n\rangle_{\rm min}$,  depends linearly on the rate of dissipation  $\gamma_d$. This is illustrated in Fig.~\ref{fig5}(b). While such behavior is common to the performance of the SS scheme, the resulting number of excitations is above the CS counterpart.

\section{Conclusions}\label{sec5}
We have presented a method to cool down a nonlinear mechanical resonator via projective measurements performed on a spin coupled to the oscillator via a Jaynes-Cummings interaction term. We have illustrated  a repeated-measurement scheme and a single-shot one. While the former requires the application of concatenated time evolutions and spin projective measurements, the single-shot scheme relies on a time-dependent tuning of the spin frequency. The time-dependent profile is designed in such a way that, after the optimized time evolution, a single projective measurement onto the ground state of the spin significantly reduces the excitations of the resonator state. The single-shot measurement scheme requires just a projective measurement and can be performed in a shorter time than its iterative counterpart, although it demands further control and tunability. We determine the shape of the spin frequency relying on the well-established chopped-random basis optimization method. The good performance of both methods is supported with numerical simulations, which allow us to attain the ground state of the nonlinear mechanical resonator to a very good approximation, even in the presence of distinct decoherence and noise sources.  Thanks to the generality of the Jaynes-Cummings model in a variety of situations, our results can be applied to different experimental platforms. 

\begin{acknowledgments}
The authors acknowledge the support by the SFI-DfE Investigator Programme (grant 15/IA/2864), the Royal Commission for the Exhibition of 1851, the H2020 Collaborative Project TEQ (Grant Agreement 766900), the Leverhulme Trust Research Project Grant UltraQuTe (grant nr. RGP-2018-266) and the Royal Society Wolfson Fellowship (RSWF/R3/183013).
\end{acknowledgments}

\appendix

\section{Approximate ground state of the deformed harmonic oscillator}\label{app:A}
The ground state of a non-linear deformed harmonic oscillator with a $x^4$ perturbation can be calculated using a first-order perturbation on $\epsilon$ as
\begin{align}
|\psi_{\rm gs}^{(1)}\rangle =|\psi_{\rm gs}^{(0)}\rangle +\sum_{k\geq 1}\frac{\bra{k} H_{\rm 1}\ket{0}}{E_{0}^{(0)}-E_{k}^{(0)}}\ket{k}
  \end{align}
where $H_1=\epsilon (a+\adag)^4/16$ and $|\psi_{\rm gs}^{(0)}\rangle =\ket{0}$ is the ground state of $H_{\rm 0}=\omega \adaga$, and $E_k^{(0)}=k \omega$ the eigenenergies. In this manner,
\begin{align}
  |\psi_{\rm gs}^{(1)}\rangle &=\frac{1}{\sqrt{1+39\epsilon^2/(512\omega^2)}}\times \nonumber \\
  &\times \left(\ket{0}-\frac{3\epsilon}{8\sqrt{2}\omega}\ket{2}-\frac{\sqrt{3}\epsilon}{16\sqrt{2}\omega}\ket{4}\right).
  \end{align}
The actual ground state $|\psi_{\rm gs}\rangle$ of $H=H_{\rm 0}+H_{\rm 1}$ can be approximated as $|\psi_{\rm gs}\rangle\approx  |\psi_{\rm gs}^{(1)}\rangle$ to first-order perturbation on $\epsilon$. For $\epsilon/\omega\lesssim 0.1$, we find infidelity $I=1-|\langle \psi_{\rm gs}|\psi_{\rm gs}^{(1)} \rangle|^2\lesssim 10^{-5}$. From the previous expression it is easy to find the approximate mean number of excitations in the ground state, which reads as
\begin{align}
\langle \psi_{\rm gs}^{(1)}|\adaga| \psi_{\rm gs}^{(1)}\rangle =\frac{21\epsilon^2}{128\omega^2\left(1+\frac{39\epsilon^2}{512\omega^2}\right)}\approx \frac{21\epsilon^2}{128\omega^2}+O(\epsilon^4/\omega^4)
  \end{align}

\section{Non-Gaussianity measure}\label{app:B}
We quantify the non-Gaussianity of a state $\rho$ by $\delta_{\rm G}[\rho]$ following~\cite{Genoni:10}. For that, we construct a reference Gaussian state $\rho_{\rm G}$ such that the first and second moments are equal to those of $\rho$. The non-Gaussianity of the state $\rho$ is then quantified as the quantum relative entropy between $\rho_{\rm G}$ and $\rho$, which for a single mode reads as
\begin{align}
\delta_{\rm G}[\rho]&=S[\rho ||\rho_{\rm G}]={\rm Tr}[ \rho \log \rho]-{\rm Tr}[ \rho \log \rho_{\rm G}]\\&=S(\rho_{\rm G})-S(\rho)=h \left(\sqrt{{\rm det[{\bf s}]}} \,\right) - S(\rho)
  \end{align}
where ${\bf s}$ is the covariance matrix, with elements $s_{jk}=1/2 \langle \{r_j,r_k \} \rangle -\langle r_j \rangle\langle r_k \rangle$, with ${\bf r}=(q,p)$ and $q=(a+\adag)/\sqrt{2}$ and $p=i(\adag-a)/\sqrt{2}$, and the function $h(x)=(x+1/2)\log(x+1/2)-(x-1/2)\log(x-1/2)$. Note that $S(\rho)=-\rho\log \rho$ is the von Neumann entropy.

\section{Optimal protocol for single-shot measurement cooling}\label{app:C}
In this article we find an optimal protocol of the spin frequency $\omega_{\rm A}(t)$ through CRAB optimization~\cite{Doria:11,Caneva:11,Caneva:11b} for linear nano-mechanical resonator ($\epsilon=0$). In this manner, the time-dependent Jaynes-Cummings model decouples in a set of Landau-Zener problems, as $H_{\rm JC}=-\omega_{\rm A}(t)/2\ket{g,0}\bra{g,0}+\oplus_{n=0}^\infty H_{\rm n}(t)$, where $H_{\rm n}(t)$ is the effective Jaynes-Cummings Hamiltonian in the subspace containing $n$ excitations which reads as
\begin{align}
H_{\rm n}(t)\!=\!\frac{\omega_{\rm A}(t)-\omega}{2}\tilde{\sigma}_z+\lambda \sqrt{n+1}\tilde{\sigma}_x
  \end{align}
where $\tilde{\sigma}_z=\ket{e,n}\bra{e,n}-\ket{g,n+1}\bra{g,n+1}$ and $\tilde{\sigma}^+=\ket{e,n}\bra{g,n+1}$, so that $\tilde{\sigma}_x=\tilde{\sigma}^++\tilde{\sigma}^-$. The protocol $\omega_{\rm A}(t)$ must be determined such that after a time $\tau$, the initial state $\ket{\phi_n(0)}=\ket{g,n+1}$ is brought to $\ket{e,n}$. Hence, the  optimization is then carried out by minimizing the cost function
\begin{align}
  \mathcal{C}=1-\frac{1}{N_c}\sum_{n=0}^{N_c-1}|\langle e,n |\phi_n(\tau) \rangle |^2
\end{align}
where $N_c-1$ is the last subspace considered in the optimization, and with respect to the $2N_\omega$ variables,  $\{A_m,B_m\}$ with $m=1,\ldots, N_\omega$. These variables $\{a_m,b_m\}$ define the protocol $\delta(t)=\omega_{\rm A}(t)-\omega$ as
\begin{align}
\delta(t)=\omega t(\tau-t)\left[\sum_{n=1}^{N_\omega}\left(a_n\cos(\omega_n t)+b_n\sin (\omega_n t)\right) \right].
  \end{align}
Here we consider fixed frequencies as $\omega_n=2\pi n/\tau$, and therefore they are not randomized as required by CRAB. We minimize $\mathcal{C}$ using the standard Nelder-Mead algorithm~\cite{Nelder:65}.

%

\end{document}